\documentclass[11pt]{article}
\usepackage[sort&compress, round, semicolon, authoryear]{natbib}
\usepackage{graphicx, lscape}
\usepackage{graphics, color}
\usepackage{epsfig}
\usepackage{epstopdf}
\usepackage{wrapfig}
\usepackage{caption}
\usepackage{subcaption}
\usepackage{amsmath, amsthm, amssymb, amscd}
\usepackage{latexsym}
\usepackage{multirow} 
\usepackage[running, switch*, displaymath, mathlines]{lineno}
\usepackage{array}
\usepackage{color}
\usepackage{rotating}
\usepackage[hyphens]{url}
\usepackage[hidelinks]{hyperref}  
\usepackage{cleveref}
\usepackage{fullpage}
\usepackage{fancyvrb}
\usepackage{pdfpages}
\usepackage{fmtcount}
\usepackage[margin=3.3cm]{geometry}
\usepackage{verbatim}
\usepackage[english]{babel}
\usepackage[utf8]{inputenc}
\usepackage{tikz} 
\usepackage{float}
\usetikzlibrary{arrows, decorations.pathmorphing, backgrounds, fit, positioning, shapes.symbols, chains}
\usetikzlibrary{decorations.markings}
\usepackage{lscape}
\usepackage{algpseudocode}
\usepackage{algorithm}
\usepackage{soul}
\usepackage{booktabs}
\usepackage{enumerate}
\usepackage{colortbl}
\long\def\comment#1{} 
\usepackage{booktabs}
\usepackage{yhmath}  
\usepackage{mathtools}
\usepackage{rotating}
\usepackage{diagbox}
\usepackage{bm}

\usepackage{mathtools}

\begin{document}

\title{Combined Quantile Forecasting for High-Dimensional Non-Gaussian Data} 

\author{
{\sc Seeun Park}~~and {\sc Hee-Seok Oh}\\
Seoul National University, Seoul 08826, Korea\\
and\\
{\sc Yaeji Lim}\\
Chung-Ang University, Seoul 06974, Korea
}
\date{\today}

\maketitle
\begin{abstract}
\noindent
This study proposes a novel method for forecasting a scalar variable based on high-dimensional predictors that is applicable to various data distributions. In the literature, one of the popular approaches for forecasting with many predictors is to use factor models. However, these traditional methods are ineffective when the data exhibit non-Gaussian characteristics such as skewness or heavy tails. In this study, we newly utilize a quantile factor model to extract quantile factors that describe specific quantiles of the data beyond the mean factor. We then build a quantile-based forecast model using the estimated quantile factors at different quantile levels as predictors. Finally, the predicted values at the various quantile levels are combined into a single forecast as a weighted average with weights determined by a Markov chain based on past trends of the target variable. The main idea of the proposed method is to incorporate a quantile approach to a forecasting method to handle non-Gaussian characteristics effectively. The performance of the proposed method is evaluated through a simulation study and real data analysis of $\text{PM}_{2.5}$ data in South Korea, where the proposed method outperforms other existing methods in most cases.
\end{abstract}


\section{Introduction} \label{sec:Introduction}
Forecasting plays a pivotal role in predicting future trends and making informed decisions in various fields. The main analytical methods used in the existing studies on forecasting consist of machine learning techniques and statistical models. In machine learning, models based on artificial neural networks (ANNs) and support vector machines (SVMs) are widely used \citep{hung2009artificial,jahn2020artificial,cao2003support}. For example, \cite{devadoss2013forecasting} used a multi-layer perception (MLP) model to forecast stock prices. For statistical analysis, various time series and multivariate models have been developed. Several studies have used ARIMA model to make forecasts  \citep{atique2019forecasting}, including \cite{zhang2018trend} that predicted monthly concentrations of $\text{PM}_{2.5}$ in Fuzhou, China. In addition, VAR model is a commonly used approach for multivariate time series forecasting \citep{lack2006forecasting}. Other multivariate models, such as multiple linear regression and the generalized additive model, have been applied in forecasting studies \citep{abuella2015solar,mathivha2020short}. Hybrid methods that combine these existing models have also been proposed in the literature \citep{wang2013arima, rahayu2017hybrid}. However, it is important to note that these methods typically deal with a relatively limited number of predictors. 

The growing availability of large panel datasets has increased the demand for advanced methods, and much work has been done to address this need to handle many predictor variables. \cite{de2008forecasting} considered Bayesian regression models that employ specific prior distributions to shrink the parameters, where the variables are aggregated or selected based on the chosen prior distribution. In addition, \cite{exterkate2016nonlinear} introduced kernel ridge regression to address the high dimensionality of predictors and the nonlinear relationship between the predictors and the target variable. Another notable framework is applying a factor model to forecasting, as proposed by \cite{stock2002forecasting}. They used the ordinary approximate factor model to extract a few common factors from many predictors and used these as explanatory variables in a linear regression model to forecast the variable of interest. This approach has gained considerable attention and is used widely, inspiring various methodologies, such as  \cite{forni2005generalized} and \cite{bernanke2005measuring}.

In this study, we consider a factor model approach for forecasting based on many predictors. Numerous datasets in various fields often exhibit non-Gaussian characteristics such as skewness and heavy tails. So, the traditional linear regression method (OLS estimator) and the ordinary factor models may not be suitable for accurate forecasting. Especially when the predictors contain a number of observations with excessively high values, the estimated factors may not effectively capture and summarize the information in these predictors. Therefore, there is a compelling need for improved models.

This paper proposes a novel method that addresses the above issues using a quantile factor model and quantile regression. The main contribution is to consider the quantiles of data, enabling it to forecast effectively across a wide range of data distributions. Instead of relying on average estimates, we utilize information at different quantile levels to enhance forecast accuracy and applicability to more general scenarios. The proposed method consists of two steps. The first step is to build a quantile-based forecast model. To do this, we use a quantile factor model to estimate quantile factors that explain the $\tau$th quantile of the high-dimensional predictors instead of the typical factors that describe the mean of the predictors. This approach is advantageous because the common factor structures of the data may differ across quantiles. We then construct a quantile regression model with the estimated quantile factors and fit the $\tau$th quantile regression at various levels. The second step is to combine the forecasted values at the various quantile levels into a single forecast as a weighted average. The weights are chosen by the transition probability of a Markov chain based on the past trends of the data. 

The rest of the paper is organized as follows. In Section \ref{sec:Background}, we provide background on our study and review existing methods for forecasting with factor models. Section \ref{sec:Quantile-based} presents the proposed method in detail. A simulation study is conducted in Section \ref{sec:Simulation study}, and real data analysis using $\text{PM}_{2.5}$ data in South Korea is discussed in Section \ref{sec:Analysis of data}. Finally, concluding remarks are provided in Section \ref{sec:Concluding remarks}. The R codes for implementing the numerical experiments in this study are available at \url{https://github.com/p-seeun/combined-quantile-forecasting}.

\section{Background} \label{sec:Background}

\subsection{Quantile regression}
Quantile regression by \cite{koenker1978regression} is a statistical technique used to estimate specific quantiles of the conditional distribution of a dependent variable $Y$ based on explanatory variables $X$. Unlike conventional linear regression, which estimates the conditional mean of $Y$ using the ordinary least squares method, quantile regression estimates the conditional quantiles of $Y$. The model assumes that the conditional quantile of $Y$ is a linear function of $X$, which can be expressed as 
\begin{equation}
\label{eq:lm}
    Q_{Y}(\tau|X) = X \beta(\tau),
\end{equation}
where $\tau$ is a specified quantile level. The estimator of $\beta(\tau)$ can be obtained by solving the following optimization problem 
\[
   \hat{\beta}_{\tau} =  \underset{\beta}{\operatorname{argmin}}\sum_{i=1}^{n} \rho_\tau(Y_i-X_i\beta)
\]
with a check function $\rho_{\tau}(u)=u(\tau-I(u<0))$. The resulting estimator, $\hat{\beta}_{\tau}$, is robust to outliers. In addition, the estimators on various quantiles allow us to understand the conditional distribution of $Y$ fully. It is beneficial when the distribution of $Y$ is skewed, heavy-tailed, and non-Gaussian, in which case the OLS estimator may not be the appropriate statistic. 

\subsection {Factor model} \label{sec:Factor model}
Factor models are one of the dimension reduction techniques. The basic form is expressed as
\begin{equation}\label{eq:fm}
    X_{t}=\Lambda f_{t}+\epsilon_{t} ,\hspace{3mm} t=1,\ldots,T,
\end{equation}
where $X_{t}=(x_{1t}, \ldots, x_{nt})'$, $\epsilon_{t}=(\epsilon_{1t},\ldots,\epsilon_{nt})'$, $f_{t}$ is a $r \times 1$ factor vector, and $\Lambda$ is a $n \times r$ loading matrix ($r<n$). In the model, $X_t$ is decomposed into a common component $\Lambda f_{t}$ and an idiosyncratic component $\epsilon_{t}$, which are orthogonal. In particular, the exact factor model assumes that the idiosyncratic terms are mutually orthogonal for all leads and lags, i.e.,
\begin{equation}\label{eq:fm_assump}
\mbox{Cov}(\epsilon_{it},\epsilon_{js})=0 ,\hspace{3mm} i \neq j \in \{1,\ldots,n\}, \hspace{3mm} t,s \in \{1,\ldots,T\}.
\end{equation}
Note that a normalization can set $Var(f_t)=I_n$. Then $\Sigma_n=\mbox{Cov}(X_t)$ may be decomposed as
\[
    \Sigma_n=\Lambda_n \Lambda_n^{\prime}+D_n,
\]
where $D_n=\mbox{Cov}(\epsilon_t)$ is a diagonal matrix due to the assumption of (\ref{eq:fm_assump}). It implies that the correlation between all observable variables is completely driven by the common factors. However, this assumption tends to be too restrictive for real-world datasets, so we need a model with a weaker assumption, which is an approximate factor model. 

Approximate factor models (AFMs) introduced by \cite{chamberlain1982arbitrage} allow the idiosyncratic components to have mild cross-correlations. Approximate factor models with $r$ factors are defined as a factor model in the form of (\ref{eq:fm}), satisfying the following condition that for all $n \in \mathbb{N}$, there exists $\Lambda_n$ such that $\Sigma_n=\mbox{Cov}(X_t)$ is decomposed as $\Sigma_n=\Lambda_n \Lambda_n^{\prime}+R_n$ where $R_n=\mbox{Cov}(\epsilon_t)$ and the eigenvalues of $R_n$ are uniformly bounded. In other words, denoting by $\lambda_i(A)$ the $i$th largest eigenvalue of $A$, $ \sup_{n}\lambda_1(R_n) < \infty$. The condition on $R_n$ limits the contribution of the idiosyncratic covariances to the total covariance of $X_t$. They also showed that under some conditions on $\Sigma_n$, the data $X_t$ can be fitted to an approximate factor model with $r$ factors, and the decomposition of $\Sigma_n$ is unique. Approximate factor models are suitable for datasets with mild cross-correlations and many variables $n$.

One way to estimate the factors is to apply the principal components problem to a factor model, which is equivalent to minimizing an objective function, 
\[
    V(F,\Lambda)=\frac{1}{nT}\sum_{t=1}^T\sum_{i=1}^n(x_{it}-\lambda_i^{\prime}f_t)^2
\]
for centered data $X=(x_{it})\in \mathbb{R}^{n \times T}$. Letting $r$ be the number of factors, the principal component estimator $\hat{\Lambda}$ can be computed as eigenvectors of $XX'$ corresponding to the $r$ largest eigenvalues, and $F=(f_1,\ldots,f_T)'$ is estimated as 
$
    \hat{F}=X'\hat{\Lambda}/n. 
$
As an extension of the approximate factor models, \cite{chen2021quantile} developed a quantile factor model (QFM), which captures factors that may differ across quantiles of data. There have been several studies in the literature that emphasize the need for quantile factors. For example, \cite{ando2020quantile} found that the factor structures explaining asset returns vary across quantiles. The QFM of the $\tau$th quantile has the form,  
\begin{equation}\label{eq:qfm}
   Q_{x_{it}}(\tau|f_{t,\tau})=\lambda^{\prime}_{i,\tau}f_{t,\tau}, 
\end{equation}
where $f_{t,\tau}$ and $\lambda_{i,\tau} \in \mathbb{R}^{r(\tau)}$ are common factors and factor loadings, respectively. That is, 
\[
    x_{it}=\lambda^{\prime}_{i,\tau}f_{t,\tau}+u_{it,\tau},
\]
where the idiosyncratic error, $u_{it,\tau}$, satisfies $P(u_{it,\tau}\leq0|f_{t,\tau})=\tau$. For the estimation of (\ref{eq:qfm}), \cite{chen2021quantile} suggested minimizing the following objective function, 
\[
    M(F,\Lambda)=\frac{1}{nT}\sum_{t=1}^T\sum_{i=1}^n \rho_{\tau}(x_{it}-\lambda^{\prime}_{i,\tau}f_{t,\tau}), 
\]
where $F=(f_{1,\tau},\ldots,f_{T,\tau})^{\prime}$ and $\Lambda=(\lambda_{1,\tau},\ldots,\lambda_{n,\tau})^{\prime}$. 
Unlike the conventional PCA method used in AFM, the objective function $M(F,\Lambda)$ is not convex in $(F,\Lambda)$, and thus, there is no closed form for the global minimum of $M(F,\Lambda)$. However, given $F=F^*$, $M(\Lambda,F^*)$ is convex in $\Lambda$, and given $\Lambda=\Lambda^*$, $M(\Lambda^*,F)$ is convex in $F$. Based on these facts, an iterative algorithm can solve the optimization problem efficiently. 

\subsection{Forecasting using factor model} \label{sec:Forecast using factor models}
In this section, we briefly describe the forecasting method proposed by \cite{stock2002forecasting}. It mainly uses approximate factor models to summarize and reduce the dimensionality of predictor variables in a large data panel. Then it uses the extracted factors to build a linear regression model for the variable of interest to forecast.

For a large number of predictors $n$ and a large number of observations $T$, let $\{X_t\}$ be a $n$-dimensional time-series data and $\{y_t\}$ be a scalar time series data that we aim to forecast. The data are observed for time points $t=1,\ldots,T$, and the goal is to build a $h$-step ahead forecast at time $T$, i.e., to forecast $y_{T+h}$.
First, the high-dimensional data $\{X_t\} \in R^n$ are summarized with latent factors $\{f_t\} \in R^r (r \ll n) $ by an approximate factor model,
\begin{equation} \label{eq:fm_t}
    X_t=\Lambda f_t+e_t
\end{equation} allowing error terms to have some mild serial and cross-sectional correlation. Next, a linear regression model is fitted between $y_{t+h}$ and ($f_t$, $y_t$) as
\begin{equation} \label{eq:frcst eqtn}
    y_{t+h}=\beta_f^{\prime}{f_t}+\beta_y^{\prime}y_t+\epsilon_{t+h}.
\end{equation}
In this model, $y_t$ is added because it is considered to affect $y_{t+h}$. Given the estimated factors and coefficients, $y_{T+h}$ can be forecasted as
\begin{equation} \label{eq:frcst eqtn_fit}
    \hat{y}_{T+h}=\hat{\beta}_f^{\prime}\hat{f}_T+\hat{\beta}_y^{\prime}y_T. 
\end{equation}
The estimation of (\ref{eq:fm_t}) can be done via the principal components method described in Section \ref{sec:Factor model}. \cite{stock2002forecasting} showed that $\hat{f}_t$ is a consistent estimator of the true latent factor subject to normalization under some assumptions that limit the strength of serial- and cross-correlation. In addition, some additional assumptions on the forecasting model in (\ref{eq:frcst eqtn}) induce the consistency of regression coefficients $\hat{\beta}_f, \hat{\beta}_y$ and allow the resulting forecast $\hat{y}_{T+h}$ to converge in probability to the optimal infeasible forecast  $\beta_f^{\prime}{f_t}+\beta_y^{\prime}y_t$, as $n,~T \rightarrow \infty$.

\section{Proposed method} \label{sec:Quantile-based}

\subsection{Quantile-based forecast method} \label{sec:Proposed method}
The main idea of the proposed method is to use quantile regression and quantile factor models for forecasting to consider the quantiles of data. This is an improvement on the forecasting method of \cite{stock2002forecasting}, which is based on mean factor estimation and ordinary least squares regression when the data follows a non-Gaussian distribution. For non-Gaussian data that are skewed or heavy-tailed, the sample mean estimated by OLS regression may not provide accurate information about the data distribution because it is easily affected by extreme observations. On the other hand, information about various quartiles of the data can better characterize the distribution and provide more accurate forecasting. 

Furthermore, given covariate $X$, integrating the quantile function of response variable $Y$ across the entire domain [0,1] yields the mean of the distribution \citep{koenker2005quantile}, that is, 
\[
    E(Y|X=x)=\int_{0}^1 Q_Y(\tau|x) d\tau 
\]
which implies that $E(Y|X=x)$ can be estimated as an average of $Q_Y(\tau|x)$ across $\tau$. In practice, we approximate the conditional mean by combining the finite number of quantile estimates 
\[
\sum_{\ell=1}^m w_\ell\hat{Q}_{Y}(\tau_\ell|x)\approx E(Y|X=x), 
\]
where $w_\ell$ denote the weight for each level $\tau_\ell$. Note that the ideal weight may be the probability of occurrence of $Q_Y(\tau|X)$ for each level $\tau$. More information on selecting the weights will be discussed later. 

For forecasting a target variable based on a large number of predictors, we propose a combined quantile-based forecast method that consists of the following two steps. In the first step, we construct a quantile-based forecast model. Given predetermined $m$ quantile levels $(\tau_1, \ldots, \tau_m)$, we make a $h$-step ahead forecast of the $\tau_\ell$th quantile of $y$ at time $T$, i.e., estimate $Q_{y_{T+h}}(\tau_\ell|X_T)$ for $\ell=1,\ldots,m$, where $y$ is the target variable to be forecast and $X$ denotes the high-dimensional predictors. By conditioning on $X_T$, it indicates that the training data at time $T$, i.e., $X_{1}, \ldots, X_{T}$, are given. To this end, we use the quantile regression model of (\ref{eq:lm}) and the quantile factor model of (\ref{eq:qfm}). For fitting $Q_{y_{t+h}}(\tau_\ell|X_t)$ by the quantile regression model, it would be more appropriate to use the $\tau_\ell$th quantile factor of $X_t$ as a predictor in the model rather than the average factor used in (\ref{eq:frcst eqtn}). In other words, expecting that $y_t$ and $X_t$ are likely to have similar behavior, it is reasonable to choose a quantile feature of $X_t$, such as the $\tau_\ell$th quantile factor, as a predictor in the quantile regression model. This step is discussed further in Section \ref{sec:Extension of}. We now extract the $\tau_\ell$th quantile factor $f_{t,\tau_\ell}$ through the quantile factor model of (\ref{eq:qfm}) for $\ell=1,2,\ldots,m$. Then, we build a quantile regression model with $y_{t+h}$ and ($\hat{f}_{t,\tau_\ell}$, $y_t$) as
\begin{equation} \label{eq:quantile forecast}
    Q_{y_{t+h}}(\tau_\ell|X_t)=\beta^{\prime}_f(\tau_\ell)\hat{f}_{t,\tau_\ell}+\beta^{\prime}_y (\tau_\ell) y_t.
\end{equation}
Here, $y_t$ is one of the most relevant data to describe $Q_{y_{t+h}}(\tau_i|X_t)$, so it is added to the forecast model as in (\ref{eq:frcst eqtn}). Hence, the $\tau_\ell$th quantile forecast is obtained by computing
\[
    \hat{Q}_{y_{T+h}}(\tau_\ell|X_T)=\hat{\beta}^{\prime}_f(\tau_\ell) \hat{f}_{T,\tau_\ell}+\hat{\beta}^{\prime}_y (\tau_\ell) y_T. 
\]

In the second step, to obtain the final forecast $\hat{y}_{T+h}$, we combine the quantile forecasted values $\hat{Q}_{y_{T+h}}(\tau_\ell|X_T)$ at different quantile levels $\tau_\ell$ ($\ell=1,\ldots,m$) into a single forecast. For this purpose, we consider a weighted average of  $\hat{Q}_{y_{T+h}}(\tau_\ell|X_T)$ as 
\begin{equation} \label{eq:combine}
    \hat{y}_{T+h}=\sum_{\ell=1}^m w_\ell \hat{Q}_{y_{T+h}}(\tau_\ell|X_T), 
\end{equation} 
where $w_\ell$ is the probability of occurrence of $Q_{y_{T+h}}(\tau_\ell|X_T)$. To determine $w_\ell$, we adopt the idea of \cite{tang2014forecasting} that uses the Markov chain. We define $S_t$, the state at time $t$, as the order of the quantile interval containing $y_t$ by comparing the observation $y_t$ to $m$ fitted quantiles $\hat{Q}_{y_t}(\tau_\ell|X_{t-h})$, $\ell=1,\ldots,m$. For example, suppose that we have three quantile levels, $\tau_1 \leq \tau_2 \leq \tau_3$. The state $S_t$ may be assigned like 
\[
    S_t = \begin{cases} 
        1, & \text{$y_{t}\leq \hat{Q}_{y_t}(\tau_1|X_{t-h})$} \\
        3, & \text{$y_{t}\geq \hat{Q}_{y_t}(\tau_3|X_{t-h})$} \\
        2, & \text{otherwise}. \\
       \end{cases}
\]
The transition probability $P_{ij}=P(S_{t+h}=j|S_t=i)$ can be empirically obtained as $\hat{P}_{ij}=\sum_t I\{S_t=i, S_{t+h}=j\}/\sum_t I\{S_t=i\}$. If the state at current time $T$ is $S_T=k$, we can approximately regard $w_\ell$, the probability of occurrence of $Q_{y_{T+h}}(\tau_\ell|X_T)$, as the probability of being in state $\ell$ at time $T+h$, and therefore, $w_\ell=\hat{P}_{k\ell}$. 

Before closing this section, we have three remarks about the proposed method. 
\begin{itemize}
\item For the estimation of $E(y_{t+h}|X_t)$, the combination of $\hat{Q}_{y_{t+h}}(\tau_i|X_t)$ is better than the OLS estimate in terms of robustness. It can be demonstrated by comparing the influential function of the OLS estimator and the $\tau$th quantile \citep{koenker2005quantile, lima2020quantile}. Without loss of generality, according to the model $E(y_{t+h}|X_t)=X_t^{\prime}\alpha$, assuming $X_t=1$, the OLS estimator of $\alpha$ would be the sample mean. To compute the influential functions of the sample mean and the $\tau$th quantile, the contaminated distribution can be defined as $F_\epsilon=\epsilon \delta_{\zeta}+(1-\epsilon)F$, where $F$ is the original distribution and $\delta_\zeta$ is a one-point distribution with the probability mass concentrated at a point $\zeta$. For an estimator $\hat{\alpha}:=\hat{\alpha}(F)$, the influential function is written as
\[
    IF_{\hat{\alpha}}(\zeta,F) = \lim_{\epsilon \rightarrow 0} \frac{\hat{\alpha}(F_\epsilon)-\hat{\alpha}(F)}{\epsilon}.
\]
For the OLS estimator $\hat{\alpha}^{OLS}$, $\hat{\alpha}^{OLS}(F_\epsilon)=\epsilon \delta_{\zeta}+(1-\epsilon)\hat{\alpha}^{OLS}(F)$, so we obtain $IF_{\hat{\alpha}^{OLS}}(\zeta,F)=\zeta-\hat{\alpha}^{OLS}(F)$. The influence function grows infinitely high as $\zeta$ increases, implying that the outlier can significantly affect the estimator $\hat{\alpha}(F)$. On the other hand, for the $\tau$th quantile estimator $\hat{\alpha}_{\tau}$, $\hat{\alpha}_{\tau}(F_\epsilon)={F_\epsilon}^{-1}(\tau)$, and thus, $IF_{\hat{\alpha}_{\tau}}(\zeta,F)=\mbox{sgn}(\zeta-\hat{\alpha}_{\tau}(F_\epsilon))/f(F_\epsilon^{-1}(\tau))$, which is bounded by $1/f(F^{-1}(\tau))$ regardless of the choice of $\zeta$. Hence, $\hat{Q}_{y_{t+h}}(\tau_\ell|X_t)$ is a robust estimator; thus, it is better than the OLS estimator in the presence of extreme observations. Therefore, the combination of quantile estimators in (\ref{eq:combine}) is better suited for forecasting than the OLS estimator in (\ref{eq:frcst eqtn_fit}). 

\item In a similar sense to choosing the $\tau_\ell$th quantile factor over the mean factor as an explanatory variable, we believe that model (\ref{eq:quantile forecast}) could be improved by including the term $y_{t,\tau_\ell}$, the $\tau_\ell$th quantile of $y_t$, instead of $y_t$. However, we found it challenging to practically implement the model because $y_{t,\tau_\ell}$ is not observed and needs to be estimated. We have explored several methods to estimate $y_{t,\tau_\ell}$ in real data analysis but have not obtained better results, so we concluded that the current form of the model is the most suitable for our purpose. 

\item It is worth mentioning why we choose the weights $w_\ell$ in (\ref{eq:combine}) as the probability of occurrence of $Q_{y_{t+h}}(\tau_i|X_t)$. Suppose we have no information about the probability. Then, it is a natural idea to simply approximate $E(y_{t+h}|X_t)=\int_{0}^1 Q_{y_{t+h}}(\tau|X_t) d\tau \approx \sum_{\ell=1}^m  \hat{Q}_{y_{t+h}}(\tau_\ell|X_t) \cdot \Delta(\tau_\ell)$. In the absence of information, the probability of occurrence of $\hat{Q}_{y_{t+h}}(\tau_\ell|X_t)$ is approximately $\Delta(\tau_\ell)$ since $P(\hat{Q}_{y_{t+h}}(\tau_\ell|X_t)<y_{t+h}<\hat{Q}_{y_{t+h}}(\tau_{\ell+1}|X_t) \, | \, X_t) \approx \tau_{\ell+1}-\tau_\ell$. However, for real datasets, it is common for the observed $y_{t+h}$ to be strongly influenced by the previous observation $y_t$ when $h$ is small enough. Considering $\text{PM}_{2.5}$ concentration data as an example, suppose that the previous observation is an extreme value that is much larger than the estimated 0.9th quantile of the distribution. Then, the following observation is likely to be large since $\text{PM}_{2.5}$ concentrations do not change much in the short term. Therefore, in this case, when forecasting $y_{t+h}$, it would be more appropriate to give heavier weight to $\hat{Q}_{y_{t+h}}(0.9|X_t)$ than to give the same weight to $\hat{Q}_{y_{t+h}}(0.1|X_t)$ and $\hat{Q}_{y_{t+h}}(0.9|X_t)$. To generalize this situation, we incorporate a Markov chain with information about the movement of $y_t$ to estimate the transition probability and consider the previous state $S_t$ defined earlier. 
\end{itemize}
 
\subsection{Extension of the proposed method} \label{sec:Extension of}
When employing a factor model, it is crucial to select predictors that are likely to explain the target variable and exhibit similar characteristics to it. In this context, it is reasonable that our method in Section \ref{sec:Proposed method} assumes that the $\tau_\ell$th quantile factor of $X_t$ can effectively explain the corresponding $\tau_\ell$th quantile of $y_{t+h}$. However, there may be cases where the available information about the predictors is limited, or it is difficult to determine their similarity with the target variable. To address these situations and introduce greater flexibility into our method, we propose an extension of the model. This extension accommodates scenarios where the predictors have substantially different distributions from the target variable. In such cases, relying solely on the $\tau_\ell$th quantile factor may not provide sufficient information in the forecasting model (\ref{eq:quantile forecast}). However, including all $\tau_1, \ldots, \tau_m$th quantile factors in the model may be inappropriate due to multicollinearity and defeat the purpose of dimensionality reduction. Instead, we consider a variable selection technique that chooses a subset of $\hat{f}_{t,\tau_1}, \ldots, \hat{f}_{t,\tau_m}$ as predictors in model (\ref{eq:quantile forecast}). 

For subset selection, we consider the group LASSO quantile regression model of \cite{sherwood2022quantile}. This approach is commonly used when the predictors are grouped together, with the objective of having the coefficients within the same group be either all zero or all non-zero. The main idea is to introduce the group LASSO penalty term into the objective function of quantile regression. The group LASSO penalty, proposed by \cite{yuan2006model}, is given as $\lambda \sum_{\ell=1}^m \|W_\ell \beta^{(\ell)}\|_2$, where $\beta^{(\ell)}$ denotes the coefficient vector corresponding to the $\ell$th group of predictors, $W_l$ denotes the penalty matrix, and $\lambda$ serves as a tuning parameter. \cite{sherwood2022quantile} set $W_\ell=d_{\ell}I_{\ell}$, where $d_\ell$ is the number of predictors in $\ell$th group. In addition, they approximated the quantile loss function by the Huber loss function to enhance computational efficiency in estimating the model. In this context, the elements within each quantile factor $\hat{f}_{t,\tau_\ell}$ are regarded as grouped predictors. By applying the group LASSO quantile regression model, we can efficiently select and incorporate the quantile factors of the chosen levels into the model. It is worth noting that these quantile factors can provide more accurate information than the mean factor, even when the relationship between predictors and the target variable is unclear. However, this extension may become unnecessary if the predictors are well-selected. A simple experiment in Section \ref{sec:Extended method} shows that the extended model performs well in specific scenarios.   

\section{Simulation study} \label{sec:Simulation study}

In this section, we conduct a simulation study to evaluate the effectiveness of the proposed method when applied to various data distributions. For each specific distribution, we generate datasets of target variable and predictors, denoted as $(y_1, \ldots, y_{T+1})^{\prime} \in \mathbb{R}^{T+1}$ and $X=(x_{it}) \in \mathbb{R}^{n \times T}$, respectively, with different combinations of $(n,T)$. We aim to forecast $y_{T+1}$ for which we possess the true value, based on $(y_1, \ldots, y_{T})^{\prime}$ and $X$. To evaluate the performance of the proposed method, we compare it with several existing methods, including time-series models and the method of \cite{stock2002forecasting}. These competing methods are summarized in Section \ref{sec:Competing methods}, and Section \ref{sec:Sim results} shows the analysis results. In addition, a simulation study for the proposed extended method is discussed in Section \ref{sec:Extended method}.

\subsection{Setup} \label{sec:Setup}
We adopt the data-generating process from \cite{chen2021quantile} and \cite{stock2002forecasting} with some adjustments. The $n$-dimensional time series of predictors $X=(x_{it}) \in \mathbb{R}^{n \times T}$ are generated as 
\begin{equation} \label{eq:DGP}
x_{it}=\lambda_{1i}f_{1t}+\lambda_{2i}f_{2t}+\lambda_{3i}f_{3t}+u_{it},
\end{equation}
where \[f_{1t}=0.8 f_{1,t-1}+ \epsilon_{1t}, \, f_{2t}=0.5 f_{2,t-1}+ \epsilon_{2t}, \, f_{3t}=0.2 f_{3,t-1}+ \epsilon_{3t}, \quad \epsilon_{1t}, \epsilon_{2t}, \epsilon_{3t} \overset{i.i.d}{\sim} N(0,1),  \] and
$\lambda_{1i}, \lambda_{2i}, \lambda_{3i} \overset{i.i.d}{\sim} N(0,1)$. The initial values of the three factors, $f_{1t}, f_{2t},$ and $f_{3t}$, are drawn from their stationary distribution which are $N(0,\frac{1}{1-0.8^2})$, $N(0,\frac{1}{1-0.5^2})$, and $N(0,\frac{1}{1-0.2^2})$, respectively. To generate various datasets, we set the distribution of $u_{it}$ in three ways:
\begin{itemize}
    \item $u_{it} \sim N(0,1)$: It represents the ideal case where the data is normally distributed and is aimed at evaluating whether the proposed method is efficient. 
    
    \item $u_{it} \sim t(2)$: It generates the data that follow $t$-distribution with 2 degrees of freedom, which is symmetric and heavy-tailed.
    
    \item $u_{it} \sim Gamma(1,5)$: It generates the gamma distributed date with shape parameter $k=1$ and scale parameter $\theta=5$, which is right-skewed and has a heavy right tail. 
\end{itemize}

Furthermore, the scalar time-series data $(y_1,\ldots,y_{T+1})$ to be forecast are generated as 
\begin{equation} \label{eq:set target}
    y_t=f_{1t}+f_{2t}+f_{3t}+e_t,
\end{equation}
where
$e_t$ follows the same distribution as $u_{it}$ in model (\ref{eq:DGP}). It has the same form as model (\ref{eq:DGP}) with all loadings $\lambda_{1t}, \lambda_{2t},$ and $\lambda_{3t}$ substituted by 1, ensuring that $X_t$ and $y_t$ are likely to exhibit similar behavior. We generate data for all combinations of $(n,T)$, where $n,T$ are selected from $\{50, 100, 150\}$. For each specific $(n,T)$, 500 simulations are performed. For the generated data, we conduct the following two experiments. 

In the first experiment, we aim to demonstrate that quantile factors estimated from model (\ref{eq:qfm}) are better suited to explain the heavy-tailed distributions than the PCA estimate of mean factors, which supports one of the ideas of our method. To this end, we investigate which estimated factor spaces better explain the true factor space by obtaining $R^2$ values between them. When analyzing the data, we assume for simplicity that the number of mean factors and the number of quantile factors for a specific quantile level are already given,  both of which are three. 

The second experiment is designed to evaluate the forecasting accuracy of our method in contrast to alternative approaches. We make one-step ahead forecasts, i.e., predict $y_{T+1}$, by these methods and compare the predicted value to the true value of $y_{T+1}$ for each method. To implement the proposed method, we select a set of quantile levels as $(\tau_1, \tau_2, \tau_3, \tau_4, \tau_5)=(0.1, 0.3, 0.5, 0.7, 0.9)$ to aggregate information from different quantile ranges of the distribution. It is also noted that for these simulated data, it is not natural to consider the Markov Chain as mentioned in Section \ref{sec:Quantile-based}. This is because each time the error $u_{it}$ is drawn independently from the past errors, and as a result, the time-series errors here do not have specific trends as in real data. Therefore, for this experiment, we choose the weight $w_{l}$ in equation (\ref{eq:combine}) to be $\Delta(\tau_\ell)$, resulting in $(w_1, w_2, w_3, w_4, w_5)=(0.1, 0.2, 0.4, 0.2, 0.1)$.

\subsection{Competing methods} \label{sec:Competing methods}
For comparison, we consider four existing methods below. 
\begin{itemize}
\item Na\"ive: It forecasts the data with the latest observation, i.e., $\hat{y}_{T+h}=y_T$.

\item AR: It uses a univariate autoregressive (AR) model of order $p$, $y_t=c+\phi_1 y_{t-1}+\cdots +\phi_p y_{t-p}+\epsilon_{t}$, where $\epsilon_{t}$ is white noise, and $\{y_t\}$ is stationary. The order $p \, (\leq 6)$ is selected by minimizing AIC.  

\item ARIMA: It uses the ARIMA($p,d,q$) model, $y^{*}_t=c+\phi_1 y^{*}_{t-1}+\cdots +\phi_p y^{*}_{t-p}+\theta_1 \epsilon_{t-1}+\cdots +\theta_p \epsilon_{t-q}+\epsilon_{t}$, where $y_t^{*}$ is a $d$ times differentiated data, and $\epsilon_t$ is an error term. For the selection of $d$, the KPSS test of \cite{Kwiatkowski1992}, one of the unit root tests, is used, which selects the order $d=1$ for most stations and periods in our data, except for some that select $d=0$. Then, $p$ and $q$ are determined by minimizing AIC. 

\item SW(2002): It is the forecasting method by \cite{stock2002forecasting}, explained in Section \ref{sec:Forecast using factor models}. Note that the number of factors $r$ is required to implement this method, as in our proposed method. In this simulation study, we assume that the value of $r$ is already known, but we must estimate it in Section \ref{sec:Analysis of data} that analyzes a real data set. Thus, we briefly describe it here. It is estimated by minimizing an information criterion $IC(r)=\ln(\Tilde{V}_r)+r\cdot g(n,T)$, where $\tilde{V}_r=\min\frac{1}{nT}\sum_{t=1}^T\sum_{i=1}^n (x_{it}-\hat{\lambda}_i^{\prime} \hat{f}_t)^2$ and $g(n,T)$ is a penalty term, which is proposed by \cite{bai2002determining}. Here $\hat{\lambda}_i$ and  $\hat{f}_t$ are extracted from the ordinary mean-based factor model, AFM. 
\end{itemize}

\subsection{Results} \label{sec:Sim results}
Table \ref{table:sim R^2} lists the average adjusted $R^2$ values of regressing each element of the true factors, $f_{i}=(f_{i1},\ldots,f_{iT})^{\prime} \in \mathbb{R}^{T}$ for $i=1,2,3$, on two kinds of estimated factors, $\hat{F}^{PCA} \in \mathbb{R}^{T \times 3}$ and $\hat{F}^{QFM} \in \mathbb{R}^{T \times 15}$, over 500 simulations.  $\hat{F}^{PCA}$ denotes the PCA estimator of $(f_1, f_2, f_3)$, and  $\hat{F}^{QFM}$ is the estimator of quantile factors derived from model (\ref{eq:qfm}). $\hat{F}^{QFM}$ contains three factor elements for each $\tau$ level within the range (0.1, 0.3, 0.5, 0.7, 0.9). When the error term $u_{it}$ follows $N(0,1)$, the adjusted $R^2$ values for $f_1$ obtained by $\hat{F}^{QFM}$ are slightly lower than those by $\hat{F}^{PCA}$. This is also true for $f_2$ and $f_3$. However, it is important to note that all of these $R^2$ values exceed 0.97, indicating that $\hat{F}^{QFM}$ explains the true factors well enough. On the other hand, when $u_{it}$ follows a heavy-tailed distribution such as $t(2)$ or $Gamma(1,5)$, the overall $R^2$ values tend to decrease in comparison to the case of $u_{it} \sim N(0,1)$, especially for $\hat{F}^{PCA}$. However, it is evident that  $\hat{F}^{QFM}$ outperforms $\hat{F}^{PCA}$ in capturing the true factor space, as it consistently yields significantly higher $R^2$ values across all $(n,T)$s. The difference is especially evident in $f_2$ and $f_3$.

Next, to compare the performance of forecasts, we consider the mean absolute error (MAE) as the evaluation metric. It is defined as, 
\begin{center}
    $MAE=\dfrac{\sum \limits_{j=1}^{N_{s}} \left |\hat{y}^{(j)}_{T+1}-y^{(j)}_{T+1}\right |}{N_{s}},\quad$
\end{center}
where $N_s=500$ denotes the number of simulations, $\hat{y}^{(j)}_{T+1}$ and $y^{(j)}_{T+1}$ are the forecasted and true values, respectively, at time $T+1$ in the $j$th simulation. Table \ref{table:sim MAE} lists the average MAE values and their standard errors over 500 simulations, corresponding to each method and $(n,T)$. For the normally distributed case, both SW(2002) and the proposed method perform better than the na\"ive and time-series models. The proposed method does not always outperform SW(2002), which may be because both the estimation of mean factors and OLS regression work well in this case, making the quantile approach less necessary. However, its performance is similar to SW(2002), so we can conclude that our method is efficient. On the other hand, when the error terms follow a $t(2)$ or $Gamma(1,5)$ distribution, SW(2002) method sometimes fails to outperform the time-series models. This may be attributed to the fact that the estimated mean factor does not provide satisfactory information for the predictors $X$, which is also supported by the results in Table 1. Furthermore, since $y$ also follows a heavy-tailed distribution, the OLS regression may not perform optimally. However, the proposed method, which incorporates the quantile approach for both factor estimation and forecasting, solves this problem by considering information from different quantiles of data. As a result, the proposed method significantly improves performance and achieves the lowest MAE in almost all cases. The only exception is when $u_{it},e_t \sim t(2)$ and $(n,T)=(100,50)$.

\begin{table}
   \caption{Average adjusted $R^2$ values of regressing each element of the true factor on the estimated mean factors and the estimated quantile factors over 500 simulations. The values are computed for each $(n,T)$ and error distribution.} 
   \small
   \centering
   \setlength{\tabcolsep}{4pt}
   {\renewcommand{\arraystretch}{0.9}
   \begin{tabular}{c|cccccccccc}
   \toprule\toprule
   ($n$,$T$) & ($50,50$) & ($50,100$) & ($50,150$) & ($100,50$) & ($100,100$) & ($100,150$) & ($150,50$) & ($150,100$) & ($150,150$)\\
   \midrule

    \multicolumn{10}{c}{$u_{it} \sim N(0,1)$} \\
    \hline
$f_1, \hat{F}^{PCA}$ & 0.990 & 0.990 & 0.991 & 0.995 & 0.995 & 0.996 & 0.997 & 0.997 & 0.997 \\ 
$f_1, \hat{F}^{QFM}$ & 0.987 & 0.988 & 0.989 & 0.994 & 0.994 & 0.995 & 0.996 & 0.996 & 0.996 \\ 
\vspace{-4mm}\\ 
$f_2, \hat{F}^{PCA}$ & 0.981 & 0.983 & 0.983 & 0.991 & 0.992 & 0.992 & 0.994 & 0.995 & 0.995 \\ 
$f_2, \hat{F}^{QFM}$ & 0.976 & 0.979 & 0.979 & 0.989 & 0.989 & 0.990 & 0.993 & 0.993 & 0.993 \\ 
\vspace{-4mm}\\ 
$f_3, \hat{F}^{PCA}$ & 0.978 & 0.978 & 0.979 & 0.989 & 0.990 & 0.990 & 0.993 & 0.993 & 0.993 \\ 
$f_3, \hat{F}^{QFM}$ & 0.973 & 0.972 & 0.973 & 0.987 & 0.987 & 0.987 & 0.991 & 0.991 & 0.991 \\ 
\hline

    \multicolumn{10}{c}{$u_{it} \sim t(2)$} \\ 
    \hline
$f_1, \hat{F}^{PCA}$ & 0.827 & 0.863 & 0.882 & 0.881 & 0.913 & 0.934 & 0.895 & 0.931 & 0.946 \\ 
$f_1, \hat{F}^{QFM}$ & 0.940 & 0.961 & 0.968 & 0.970 & 0.977 & 0.985 & 0.975 & 0.986 & 0.989 \\ 
\vspace{-4mm}\\ 
$f_2, \hat{F}^{PCA}$ & 0.618 & 0.614 & 0.622 & 0.677 & 0.674 & 0.672 & 0.699 & 0.686 & 0.709 \\ 
$f_2, \hat{F}^{QFM}$ & 0.857 & 0.871 & 0.883 & 0.908 & 0.908 & 0.910 & 0.914 & 0.916 & 0.936 \\ 
\vspace{-4mm}\\ 
$f_3, \hat{F}^{PCA}$ & 0.463 & 0.452 & 0.466 & 0.501 & 0.483 & 0.447 & 0.544 & 0.497 & 0.519 \\ 
$f_3, \hat{F}^{QFM}$ & 0.775 & 0.777 & 0.797 & 0.831 & 0.817 & 0.784 & 0.856 & 0.824 & 0.847 \\
\hline

    \multicolumn{10}{c}{$u_{it} \sim Gamma(1,5)$} \\
    \hline
$f_1, \hat{F}^{PCA}$ & 0.703 & 0.785 & 0.809 & 0.841 & 0.886 & 0.897 & 0.887 & 0.924 & 0.932 \\ 
$f_1, \hat{F}^{QFM}$ & 0.871 & 0.911 & 0.924 & 0.936 & 0.960 & 0.967 & 0.954 & 0.976 & 0.979 \\ 
\vspace{-4mm}\\ 
$f_2, \hat{F}^{PCA}$ & 0.498 & 0.611 & 0.658 & 0.697 & 0.783 & 0.801 & 0.796 & 0.851 & 0.863 \\ 
$f_2, \hat{F}^{QFM}$ & 0.719 & 0.780 & 0.806 & 0.841 & 0.885 & 0.889 & 0.900 & 0.922 & 0.923 \\ 
\vspace{-4mm}\\ 
$f_3, \hat{F}^{PCA}$ & 0.407 & 0.516 & 0.568 & 0.617 & 0.734 & 0.757 & 0.728 & 0.812 & 0.830 \\ 
$f_3, \hat{F}^{QFM}$ & 0.623 & 0.680 & 0.693 & 0.771 & 0.818 & 0.821 & 0.840 & 0.868 & 0.870 \\ 
\hline
   \bottomrule
   \end{tabular}
   }
\label{table:sim R^2}
\end{table}

\begin{table}
   \caption{Average MAE values and their standard errors over 500 simulations. These are the results of one-step ahead forecasts by four competing methods and the proposed method. The values are computed for each $(n,T)$ and error distribution.} 
   \footnotesize
   \centering
   {\renewcommand{\arraystretch}{0.87}
   \setlength{\tabcolsep}{4pt}
   \begin{tabular}{c|cccccccccc}
   \toprule
   \toprule
   ($n$,$T$) & ($50,50$)& ($50,100$)& ($50,150$)& ($100,50$)& ($100,100$)& ($100,150$)&($150,50$)& ($150,100$)& ($150,150$)\\
   \midrule
   \multicolumn{10}{c}{$u_{it},e_t \sim N(0,1)$} \\
   \hline
\multirow{2}*{Na\"ive} & 2.00 & 2.02 & 2.01 & 1.98 & 1.95 & 2.03 & 1.93 & 2.01 & 2.04 \\ 
  & (0.07) & (0.06) & (0.07) & (0.07) & (0.07) & (0.07) & (0.06) & (0.07) & (0.07) \\ 
\cline{1-10} 
\multirow{2}*{AR} & 1.79 & 1.76 & 1.69 & 1.74 & 1.75 & 1.79 & 1.73 & 1.81 & 1.72 \\ 
  & (0.06) & (0.06) & (0.06) & (0.06) & (0.06) & (0.06) & (0.06) & (0.06) & (0.06) \\ 
\cline{1-10} 
\multirow{2}*{ARIMA} & 1.77 & 1.75 & 1.70 & 1.77 & 1.77 & 1.78 & 1.74 & 1.82 & 1.71 \\ 
  & (0.06) & (0.06) & (0.06) & (0.06) & (0.06) & (0.06) & (0.06) & (0.06) & (0.06) \\ 
\cline{1-10} 
\multirow{2}*{SW(2002)} & 1.74 & 1.61 & \textbf{1.60} & \textbf{1.62} & \textbf{1.68} & \textbf{1.64} & \textbf{1.63} & \textbf{1.66} & \textbf{1.56} \\ 
  & (0.06) & (0.06) & \textbf{(0.06)} & \textbf{(0.06)} & \textbf{(0.06)} & \textbf{(0.06)} & \textbf{(0.06)} & \textbf{(0.06)} & \textbf{(0.06)} \\ 
\cline{1-10} 
\multirow{2}*{Proposed} & \textbf{1.74} & \textbf{1.61} & 1.61 & 1.63 & 1.69 & 1.66 & 1.64 & 1.67 & 1.56 \\ 
  & \textbf{(0.06)} & \textbf{(0.06)} & (0.06) & (0.06) & (0.06) & (0.06) & (0.06) & (0.06) & (0.06) \\ 
 \hline
 \multicolumn{10}{c}{$u_{it},e_t \sim t(2)$} \\
   \hline
\multirow{2}*{Na\"ive} & 2.63 & 3.06 & 2.86 & 2.65 & 2.91 & 2.69 & 2.48 & 2.64 & 2.92 \\ 
  & (0.12) & (0.17) & (0.16) & (0.10) & (0.15) & (0.12) & (0.13) & (0.11) & (0.15) \\ 
\cline{1-10} 
\multirow{2}*{AR} & 2.36 & 2.52 & 2.36 & 2.31 & 2.43 & 2.25 & 2.24 & 2.33 & 2.42 \\ 
  & (0.09) & (0.13) & (0.12) & (0.08) & (0.11) & (0.09) & (0.11) & (0.09) & (0.13) \\ 
\cline{1-10} 
\multirow{2}*{ARIMA} & 2.39 & 2.50 & 2.38 & \textbf{2.28} & 2.43 & 2.30 & 2.18 & 2.33 & 2.41 \\ 
  & (0.10) & (0.12) & (0.12) & \textbf{(0.08)} & (0.11) & (0.09) & (0.11) & (0.09) & (0.13) \\ 
\cline{1-10} 
\multirow{2}*{SW(2002)} & 2.33 & 2.44 & 2.26 & 2.51 & 2.43 & 2.23 & 2.34 & 2.31 & 2.36 \\ 
  & (0.09) & (0.13) & (0.12) & (0.08) & (0.11) & (0.09) & (0.11) & (0.09) & (0.13) \\ 
\cline{1-10} 
\multirow{2}*{Proposed} & \textbf{2.22} & \textbf{2.35} & \textbf{2.14} & 2.28 & \textbf{2.34} & \textbf{2.10} & \textbf{2.15} & \textbf{2.02} & \textbf{2.30} \\ 
  & \textbf{(0.09)} & \textbf{(0.13)} & \textbf{(0.12)} & (0.08) & \textbf{(0.11)} & \textbf{(0.09)} & \textbf{(0.11)} & \textbf{(0.09)} & \textbf{(0.13)} \\ 
 \hline
\multicolumn{10}{c}{$u_{it},e_t \sim Gamma(1,5)$} \\
   \hline
\multirow{2}*{Na\"ive} & 5.27 & 5.42 & 5.04 & 5.18 & 5.31 & 5.23 & 5.38 & 5.35 & 5.28 \\ 
  & (0.22) & (0.22) & (0.21) & (0.20) & (0.24) & (0.20) & (0.23) & (0.22) & (0.25) \\ 
\cline{1-10} 
\multirow{2}*{AR} & 4.42 & 4.25 & 4.11 & 4.20 & 4.13 & 4.20 & 4.37 & 4.17 & 4.19 \\ 
  & (0.18) & (0.16) & (0.16) & (0.16) & (0.17) & (0.15) & (0.19) & (0.18) & (0.19) \\ 
\cline{1-10} 
\multirow{2}*{ARIMA} & 4.45 & 4.19 & 4.15 & 4.14 & 4.07 & 4.17 & 4.34 & 4.23 & 4.24 \\ 
  & (0.18) & (0.16) & (0.16) & (0.16) & (0.17) & (0.15) & (0.18) & (0.18) & (0.19) \\ 
\cline{1-10} 
\multirow{2}*{SW(2002)} & 4.41 & 4.09 & 3.95 & 4.20 & 4.09 & 4.02 & 4.36 & 4.10 & 4.09 \\ 
  & (0.18) & (0.16) & (0.16) & (0.16) & (0.17) & (0.15) & (0.19) & (0.18) & (0.19) \\ 
\cline{1-10} 
\multirow{2}*{Proposed} & \textbf{4.27} & \textbf{4.01} & \textbf{3.79} & \textbf{4.12} & \textbf{4.02} & \textbf{3.94} & \textbf{4.25} & \textbf{4.05} & \textbf{3.94} \\ 
  & \textbf{(0.18)} & \textbf{(0.16)} & \textbf{(0.16)} & \textbf{(0.16)} & \textbf{(0.17)} & \textbf{(0.15)} & \textbf{(0.19)} & \textbf{(0.18)} & \textbf{(0.19)} \\ 
 \hline
   \bottomrule
   \end{tabular}
   }
\label{table:sim MAE}   
\end{table}

\subsection{Performance of the proposed extended method} \label{sec:Extended method}
In this section, we conduct an experiment using the proposed extended method under specific circumstances where some predictors do not have a similar distribution to the target variable. 

For this purpose, We use the same setup as in Section \ref{sec:Setup}, except for the distribution of $u_{it}$, which represents the error term used when generating the predictors. We randomly divide the numbers from 1 to $n$ into two equal sets, labeling them A and B. Then we set $u_{it} \sim t(2)$ for $i \in A$ and $u_{it} \sim Gamma(1,5)$ for $i \in B$. Note that $t(2)$ is symmetric with heavy tails on both sides, and $Gamma(1,5)$ is right-skewed with a heavy right tail. As a result, this approach models a situation in which half of the predictor variables exhibit different behavior from the target variable. Instead of using $\hat{f}_{t,\tau_\ell}$ as predictors to forecast $Q_{y_{t+h}}(\tau_i|X_t)$ in model (\ref{eq:quantile forecast}), we choose a subset from the quantile factors on all levels through the group LASSO quantile regression, as described in Section \ref{sec:Extended method}. We use the R package \textbf{hrqglas}, which implements the estimation method by \cite{sherwood2022quantile}. We then compare the performance of the extended method with the proposed method and the four existing methods through the MAE values. We repeat this process for several combinations of $(n,T)$s.  

The simulation results are listed in Table \ref{table:ext MAE}. In most cases, SW(2020) performs poorly or similarly compared to the time-series models despite utilizing information from the predictors, while the proposed method performs better than the time-series models in most cases. The extended method performed even better than the proposed method, yielding the lowest MAE values in most cases. The only exception is when  $(n,T)=(50,150)$, which is almost identical to the best result. This leads to the conclusion that the proposed method is improved in this scenario by adding a variable selection step into equation (\ref{eq:quantile forecast}). 



\begin{table}[ht]
   \caption{Average MAE values and their standard errors over 500 simulations. These are the results of one-step ahead forecasts by four competing methods, the proposed method, and the extended method. The values are computed for each $(n,T)$ when $u_{it} \sim t(2)$ and $e_t \sim Gamma(1,5)$.} 
   \footnotesize
   \centering
   {\renewcommand{\arraystretch}{0.87}
   \setlength{\tabcolsep}{4pt}
   \begin{tabular}{c|cccccccccc}
   \toprule
   ($n$,$T$) & ($50,50$)& ($50,100$)& ($50,150$)& ($100,50$)& ($100,100$)& ($100,150$)&($150,50$)& ($150,100$)& ($150,150$)\\
    \midrule
    \hline
\multirow{2}*{Na"ive} & 5.41 & 5.26 & 5.52 & 5.51 & 5.36 & 5.26 & 5.95 & 4.83 & 5.61 \\ 
  & (0.23) & (0.21) & (0.25) & (0.23) & (0.23) & (0.21) & (0.26) & (0.19) & (0.23) \\ 
\cline{1-10} 
\multirow{2}*{AR} & 4.28 & 4.14 & 4.25 & 4.38 & 4.10 & 4.01 & 4.30 & 4.09 & 4.45 \\ 
  & (0.17) & (0.17) & (0.19) & (0.18) & (0.17) & (0.16) & (0.18) & (0.15) & (0.18) \\ 
\cline{1-10} 
\multirow{2}*{ARIMA} & 4.29 & 4.10 & 4.26 & 4.32 & 4.11 & 4.00 & 4.26 & 4.04 & 4.47 \\ 
  & (0.17) & (0.16) & (0.19) & (0.18) & (0.17) & (0.16) & (0.18) & (0.15) & (0.18) \\ 
\cline{1-10} 
\multirow{2}*{SW(2002)} & 4.26 & 4.16 & 4.16 & 4.69 & 4.10 & 3.93 & 4.39 & 4.02 & 4.98 \\ 
  & (0.17) & (0.16) & (0.18) & (0.27) & (0.19) & (0.16) & (0.21) & (0.15) & (0.58) \\ 
\cline{1-10} 
\multirow{2}*{Proposed} & 4.01 & 4.02 & \textbf{4.00} & 4.34 & 3.90 & 3.81 & 4.18 & 3.89 & 4.51 \\ 
  & (0.17) & (0.17) & \textbf{(0.19)} & (0.24) & (0.18) & (0.16) & (0.19) & (0.15) & (0.36) \\ 
\cline{1-10} 
\multirow{2}*{Extended} & \textbf{3.99} & \textbf{4.00} & 4.00 & \textbf{4.19} & \textbf{3.83} & \textbf{3.81} & \textbf{4.02} & \textbf{3.82} & \textbf{4.28} \\ 
  & \textbf{(0.17)} & \textbf{(0.17)} & (0.19) & \textbf{(0.19)} & \textbf{(0.17)} & \textbf{(0.16)} & \textbf{(0.18)} & \textbf{(0.15)} & \textbf{(0.20)} \\ 
 \hline
   \bottomrule
   \end{tabular}
   }
\label{table:ext MAE}   
\end{table}

\section{Analysis of South Korea $\text{PM}_{2.5}$ data} \label{sec:Analysis of data}

In this section, we perform a real data analysis using the proposed method. The dataset consists of hourly $\text{PM}_{2.5}$ measurements observed at 308 stations in South Korea. Let $\{y_t\}$ be time series data of $\text{PM}_{2.5}$ concentration observed at a particular station, and $X_t$ be multivariate time series data observed at the others. For each test period and each $h=1,2,\ldots,6$, we forecast $y_{T+h}$ at every time $T$ such that $T+h$ is included in the test period. For each forecasting, the training data are the observations for the past 500 hours, which are $T-499, \ldots, T$. So, for the chosen station, we can compute $\hat{y}_{T+h}$ and repeat this process for each station to make complete forecasting. To evaluate the performance of the proposed method, we compare it to the four existing methods in Section \ref{sec:Simulation study} and an additional method specifically suited to this $\text{PM}_{2.5}$ data. Section \ref{sec:Data description} provides a detailed description of the data we use. Section \ref{sec:Data analysis} outlines each step of the processed analysis, and the results are presented in Section \ref{sec:PM results}.

\subsection{Data description} \label{sec:Data description}
\begin{figure}[!ht]
        \centering
        \includegraphics[height=9cm]{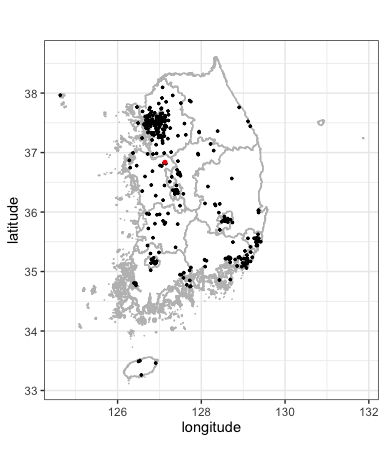} \vspace{-12mm} \caption*{(a)} 
        \includegraphics[height=4cm]{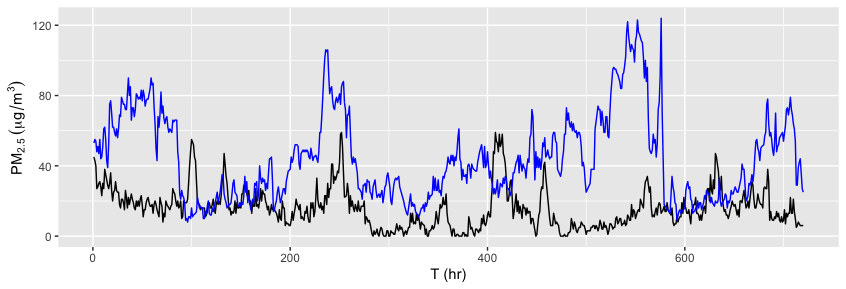} \vspace{-5mm} \caption*{(b)} 
        \vspace{-5mm}
        \caption{(a) Distribution of monitoring stations. (b) Time series plots of $\text{PM}_{2.5}$ observed at the stations marked in red in panel (a). The black and blue lines represent 30-day plots from August 23, 2019, and January 23, 2020, respectively.}       
        \label{fig:station_map}
\end{figure}

The data we use are hourly $\text{PM}_{2.5}$ data observed for 8592 hours from May 24, 2019, to May 15, 2020, at 308 stations across South Korea. The original data are provided by Air Korea (\url{www.airkorea.or.kr}), operated by the Korean Ministry of Environment. The location of stations is shown in Figure \ref{fig:station_map}, which shows that they are densely placed around large cities, especially Seoul, the capital city located in the northwest area. 
For example, the time series data measured at the station marked in red are shown in the bottom panel of Figure \ref{fig:station_map}. The black and blue lines denote the time series plots for 30 days from August 23, 2019, and January 23, 2020, respectively. As shown in the figure, the $\text{PM}_{2.5}$ concentration varies over time and sometimes has very high values. 

In this analysis, we aim to forecast $h$-hour in advance ($h=1,2,\ldots,6$) from a given time point $T$ for all stations. For each $h$, this process is repeated for all $T$ such that $T+h$ is included in 12 periods of 48 hours each, one month apart (period 1: June 14--15, 2019, period 2: July 14--15, 2019, $\ldots$ , period 12: May 14--15, 2020). Since environmental data in South Korea are highly affected by the seasons, we chose 12 time periods, each one month apart, to avoid getting only good predictions by chance. Using this dataset, we want to predict the $\text{PM}_{2.5}$ concentration at each station, taking the remaining stations as predictor variables. 

\begin{figure}
        \centering
        \includegraphics[height=6cm]{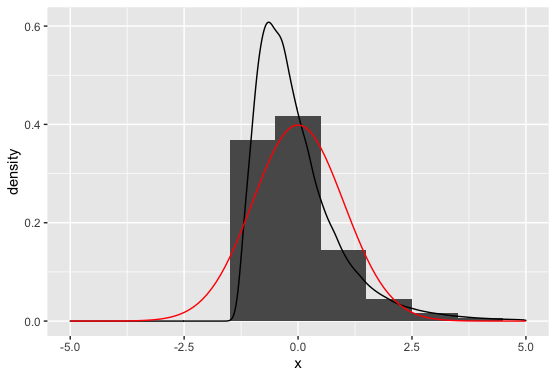}
         \vspace{-5mm}
        \caption{Histogram of standardized $\text{PM}_{2.5}$ data in South Korea and its kernel density estimate (black line). The red line denotes the density function of the standard normal distribution.}
        \label{fig:data_hist}
\end{figure}

One of the notable things about the $\text{PM}_{2.5}$ data we want to analyze is that it exhibits non-Gaussian characteristics, as shown in Figure \ref{fig:data_hist}. Compared to the standard normal distribution of the red line, the data distribution is skewed to the right and has a heavy right tail, meaning that there are quite a few observations with high $\text{PM}_{2.5}$ concentrations. This deviation of the data from the normal distribution is also supported by the high kurtosis of 7.6.

Furthermore, because air quality monitoring systems undergo several screening processes to remove abnormal data, the data at each station often contain missing values. In this analysis, we have previously deleted the stations with a missing rate of more than 20\% and used only the remaining 308 stations as data. As a result, there are 5.4\% of missing data in total, and to make it complete, we impute missing values with estimators based on the observed data. For imputation, we employ the EM algorithm proposed by \cite{stock2002forecasting}. Let $x_{it}$ be the $\text{PM}_{2.5}$ concentration observed at time $t$ and station $i$, and $X=(x_{it})_{1 \leq i \leq n, 1\leq t \leq T}$ be the full data. Given the number of factors $r$,  the detailed imputation steps are as follows.  

\begin{algorithm}[H] \vspace{1mm} 
\begin{algorithmic}[1]
\State For missing values $x_{it}$, impute initial values based on observations at near stations and close time points.
\State Update $\hat{\Lambda}$ as multiplying $\sqrt{n}$ to $r$ eigenvectors of $XX'$ corresponding to the $r$ largest eigenvalues and $\hat{F}$ as $X^{\prime}\hat{\Lambda}/n$.
\State Compute $x_{it}=\hat{\lambda}_i^{\prime} \hat{F_t}$.
\State Repeat 2--3 until convergence.
\end{algorithmic}
\caption{EM algorithm for imputation of missing data}
\label{alg:seq}
\end{algorithm}

\subsection{Data analysis} \label{sec:Data analysis}
We choose the quantile levels as $(\tau_1, \tau_2, \tau_3, \tau_4, \tau_5)=(0.1, 0.3, 0.5, 0.7, 0.9)$, consistent with the simulation study. To fit the quantile factor model, the number of factors $r(\tau_\ell)$ must be set for each $\tau_\ell$. For each test period, $r(\tau_\ell)$ is determined based on the information criteria proposed by \cite{ando2020quantile}, 
\[
IC(r)=\ln(\hat{V}_r(\tau))+r\cdot g(n,T),
\]
where $\hat{V}_r(\tau)=\min\frac{1}{nT}\sum_{t=1}^T\sum_{i=1}^n \rho_\tau(x_{it}-\hat{\lambda}_{i,\tau}^{\prime} \hat{f}_{t,\tau})$ and $\hat\lambda_{i,\tau}$, $\hat{f}_{t,\tau}$ are obtained by fitting the $\tau$ quantile factor model with $r$ factors. The $g(n,T)$ is a penalty term to prevent overfitting when $r$ gets too large. In this study, we use $g(n,T)=(\frac{n+T}{nT})\ln(\frac{nT}{n+T})$. Letting $X^{c}= (x_{it}) \in \mathbb{R}^{308 \times 500}$ be the total data observed in the training period,
$\hat{r}(\tau)$ is obtained by 
\begin{equation} \label{eq:IC_QFM}
    \hat{r}(\tau)=\underset{r}{\operatorname{argmin}}~ IC(r). 
\end{equation}
In the forecasting problem, we should use the 307 predictors data, $X$, rather than the complete training data, $X^c$. The $X$ depends on the station we choose to make a forecast on. So,  we approximate all $X$'s to the complete training data, $X^c$, for simplicity. In addition, the training data $X^c$ slightly differs across all time points in the forecast period. However, we observed that the difference in the selected $r(\tau_\ell)$ across the time points in the period was minor. Therefore, we choose only one $r(\tau_\ell)$ for each period based on the average value of $r(\tau_\ell)$ across several time points in the period. 
Table \ref{table:numoffac_qfm} lists the selected $r(\tau_\ell)$, $\ell=1,\ldots,5$ for all test periods. 


\begin{table}[h] 
\centering
\def\arraystretch{1.0}
\caption{The selected number of factors for each quantile level and test period.}
\vspace{-3mm}
\begin{tabular}{|c|cccccccccccc|}
\cline{2-13}
\multicolumn{1}{c}{} & \multicolumn{12}{|c|}{Period}  \\
\hline
$\tau$ & {1} & {2} & {3}& {4} & {5} & {6} & {7} & {8} & {9}& {10} & {11} & {12} \\
\hline
0.1 & 10 & 10 & 9 & 8 & 10 & 9 & 8 & 10 & 12 & 10 & 8 & 9  \\
0.3 & 9 & 9 & 7 & 7 & 9 & 9 & 8 & 8 & 10 & 10 & 7 & 8 \\
0.5 & 9 & 9 & 7 & 7 & 9 & 8 & 8 & 8 & 10 & 10 & 8 & 8 \\
0.7 & 10 & 10 & 7 & 9 & 10 & 8 & 8 & 8 & 11 & 10 & 9 & 9  \\
0.9 & 13 & 12 & 11 & 11 & 13 & 9 & 8 & 12 & 13 & 12 & 9 & 11 \\
\hline
\end{tabular}
\label{table:numoffac_qfm}
\end{table}

To estimate the Markov chain probabilities, the state $S_t$ is obtained by the procedure described in Section \ref{sec:Quantile-based} as 
\[
    S_t = \begin{cases} 
        1, & \text{$y_{t} \leq \hat{Q}_{y_t}(0.1|X_{t-h})$} \\
        2, & \text{$\hat{Q}_{y_t}(0.1|X_{t-h}) \leq y_{t}\leq \hat{Q}_{y_t}(0.3|X_{t-h})$} \\
        3, & \text{$\hat{Q}_{y_t}(0.3|X_{t-h}) \leq y_{t}\leq \hat{Q}_{y_t}(0.7|X_{t-h})$} \\
        4, & \text{$\hat{Q}_{y_t}(0.7|X_{t-h}) \leq y_{t}\leq \hat{Q}_{y_t}(0.9|X_{t-h})$} \\
        5, & \text{$\hat{Q}_{y_t}(0.9|X_{t-h}) \leq y_{t}$}. \\
       \end{cases}
\]
Then, for each station in each period, the transition probability $P_{ij}=P(S_{t+h}=j|S_t=i)$ is empirically estimated based on the data observed at the 10 nearest stations, including itself, during the corresponding period. 

\begin{figure} 
\centering
\hspace{-2mm}
\begin{subfigure}{0.33\textwidth}
    \centering
    \includegraphics[width=4.9cm]{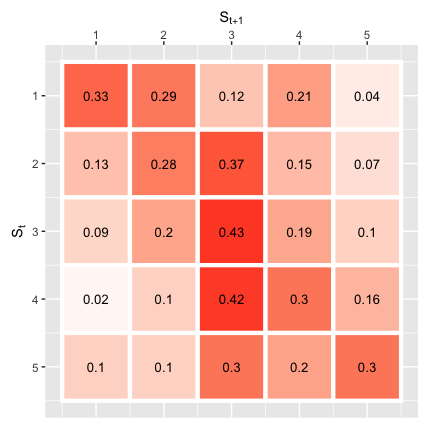}  
    \caption{Period 1.}
    \label{subfig:a}
\end{subfigure} \hspace{-6mm}
\begin{subfigure}{0.33\textwidth}
    \centering
    \includegraphics[width=4.9cm]{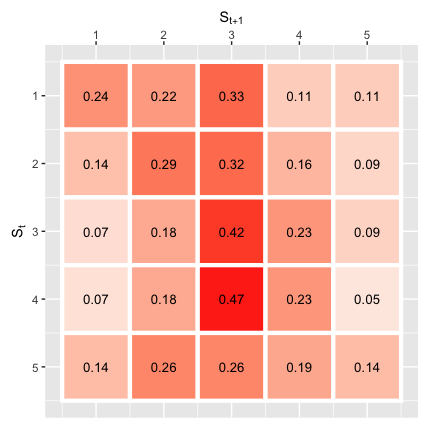}  
    \caption{Period 5.}
\end{subfigure} \hspace{-6mm}
\begin{subfigure}{0.33\textwidth}
    \centering
    \includegraphics[width=4.9cm]{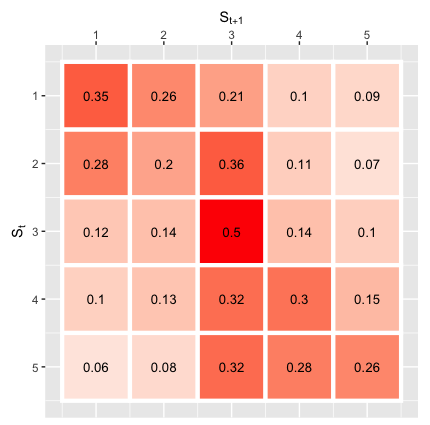}  
    \caption{Period 9.}
\end{subfigure} \hspace{-2mm}
\begin{subfigure}{0.05\textwidth}
    \centering
    \includegraphics[height=4.8cm]{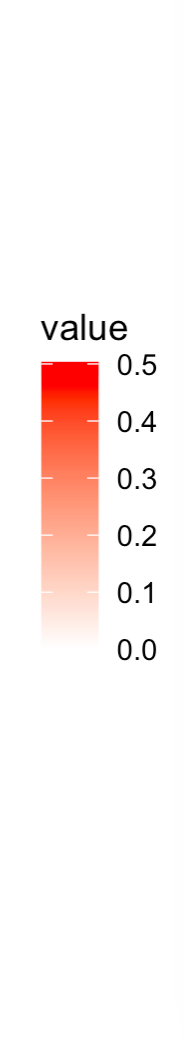}%
    \caption*{ }
\end{subfigure}
\caption{Heatmap of estimated transition probability matrix corresponding to the station marked in red in Figure \ref{fig:station_map}, for three periods when $h=1$. The element of $i$th row and $j$th column represents the transition probability $P_{ij}=P(S_{t+h}=j|S_t=i)$. If $S_T=i$, the weights $(w_1, w_2, w_3, w_4, w_5)$ on time $T$ are determined as the values of the $i$th row.}
\label{fig:heatmap}
\end{figure}

Figure \ref{fig:heatmap} illustrates the transition probability matrices corresponding to the station colored in red in Figure \ref{fig:station_map}. These matrices are obtained for periods 1, 5, and 9 with $h=1$. The rows represent $S_t$, the current state, while the columns represent $S_{t+1}$, the state one hour later. Each element in the $i$th row and $j$th column corresponds to the transition probability $P_{ij}=P(S_{t+h}=j|S_t=i)$. For example, if $S_T = 1$ for time $T$ in period 1, the weights used to predict $y_{T+1}$ would be $(0.33, 0.29, 0.12, 0.21, 0.04)$, which is the first row of the matrix in Figure \ref{subfig:a}. This is markedly different from the weights $(0.1, 0.2, 0.4, 0.2, 0.1)$, which are estimated without transition probability information. In particular, the probability at the lower quantile levels after one hour, i.e., $S_{T+1}=1$ or $S_{T+1}=2$, has much higher values than 0.1 or 0.2, which is consistent with the expectation that $\text{PM}_{2.5}$ concentration does not change dramatically in a short time. In general, all three matrices consistently show that the five diagonal elements have values higher than 0.1, 0.2, 0.4, 0.2, and 0.1, respectively. Furthermore, the elements close to the diagonal tend to have significantly higher values than the weights based solely on interval lengths, as opposed to those farther away from the diagonal. These observations indicate that the past $\text{PM}_{2.5}$ movement has had a noticeable impact on the weights. 

The four existing methods used in Section \ref{sec:Simulation study} are also applied to the data to compare the performance. For time-series models, when $h>1$, the forecasting procedure is performed recursively, meaning that the forecasted value $\hat{y}_{t+1}$ is used as an observation to forecast $y_{t+2}$. Then $\hat{y}_{t+2}$ is used to forecast $y_{t+3}$, and so on until $\hat{y}_{t+h}$ is obtained. We also apply an additional competing method called ``Near-stations" below. For forecasting, it fits the multiple regression model, $y_{t+h}=\beta_f^{\prime}{X^*_t}+\beta_y^{\prime}y_t+\epsilon_{t+h}$, where $\{X^*_t\} \in \mathbb{R}^r$ denote the observed data at time $t$ from the $r$ nearest stations. The number of stations, $r$, is selected to match the number of factors used in SW(2002) described in Section \ref{sec:Competing methods}. 

Before closing this section, it is worth noting that the extended method is not required for this real-world data analysis since both $y_t$ and $X_t$ represent $\text{PM}_{2.5}$ concentrations. This setting satisfies the condition that the quantile features of the predictor variable effectively describe the same quantile of the target variable, so the proposed method is sufficient. 

\subsection{Results} \label{sec:PM results}
We performed $h$-step ahead forecasts $(h = 1,\ldots,6)$ over 12 test periods using the proposed method, the extended method, and the five other approaches. We compute the mean absolute error (MAE) of forecasts to evaluate the performance as
\begin{center}
    $MAE=\dfrac{\sum \limits_{i=1}^{n}\sum \limits_{t=t_1}^{t_T} \left |\hat{y}_{t,i}-y_{t,i}\right |}{nT},\quad (n=308,~T=48)$,
\end{center}
where $t_1, \ldots, t_T$ denote the time points in a test period, $\hat{y}_{t,i}$ and $y_{t,i}$ are the forecasted and true values, respectively, at time $t$ and station $i$. Note that we consider 308 stations and a test period of 48 hours for forecasting.

Tables \ref{table:result1} and \ref{table:result2} list MAE values for each case depending on $h$, test period, and method. Overall, the proposed method or the extended method outperforms the other methods in most cases, except for some cases of periods 9 and 10. Even for periods  9 and 10, the proposed method shows results very similar to the best results. The fact that the results are consistently more accurate when using SW(2002) and Near-stations model than AR and ARIMA models indicates that incorporating information from other stations improves forecasts. There are many cases where SW(2002) method improves the forecast compared to the Near-stations model, such as in periods 6, 7, 8, 10, 11, and 12. The results suggest that in these cases, the factors contain better information than observations from the same number of nearest stations. This implies that the factors summarize the data $X_t$ better than the selected predictors based on the distance between them. 

On the other hand, there are still some cases where the factor model does not improve the results over the Near-stations model. However, by incorporating the quantile approach, the proposed method leads to more accurate forecasts and outperforms the other methods in most cases. We have observed that the quantile approach did not outperform SW(2020) for some cases in periods 8 and 10. For example, the $\text{PM}_{2.5}$ concentration in period 8 exhibits a relatively normal distribution with low skewness and kurtosis compared to other test periods, where SW(2020) performs better. Nevertheless, it is worth noting that the difference between the MAE values of our method and SW(2020) is insignificant. This implies that the proposed method is efficient and performs well, although it does not perform best in some cases. 

Furthermore, we examined the extended version of the proposed method. In general, the extended method performed similarly to the proposed method, with no consistent pattern of one method outperforming the other. However, at times, it can achieve significantly better results. This is particularly noticeable in periods 7, 8, and 11, where significant differences in MAE values are observed. 

\begin{figure}
\centering
\begin{subfigure}{0.32\textwidth}
    \centering
    \includegraphics[width=5cm]{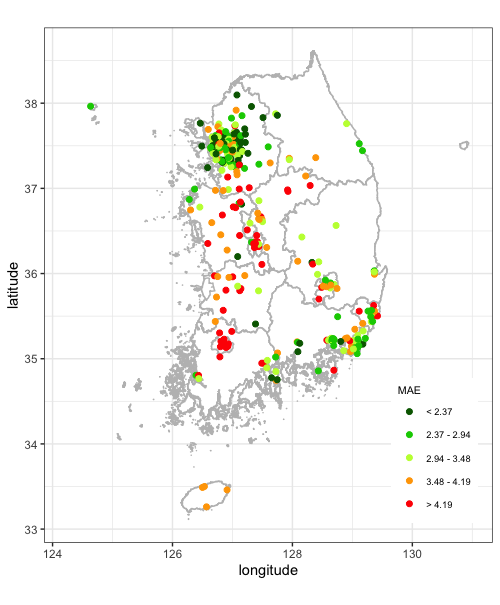}  
    \caption{Period 1, $h=1$.}
\end{subfigure}
\begin{subfigure}{0.32\textwidth}
    \centering
    \includegraphics[width=5cm]{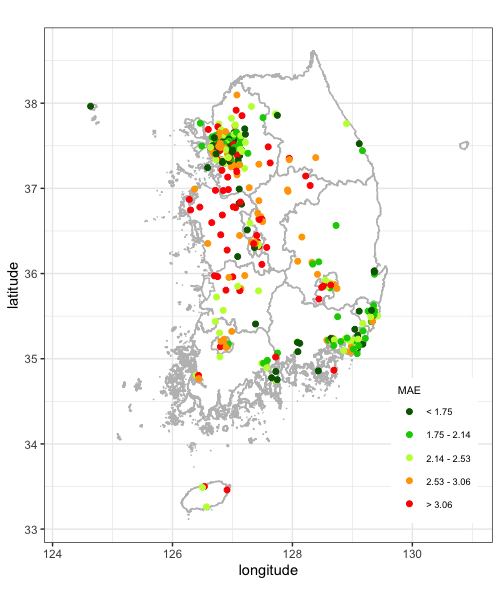}  
    \caption{Period 5, $h=1$.}
\end{subfigure}
\begin{subfigure}{0.32\textwidth}
    \centering
    \includegraphics[width=5cm]{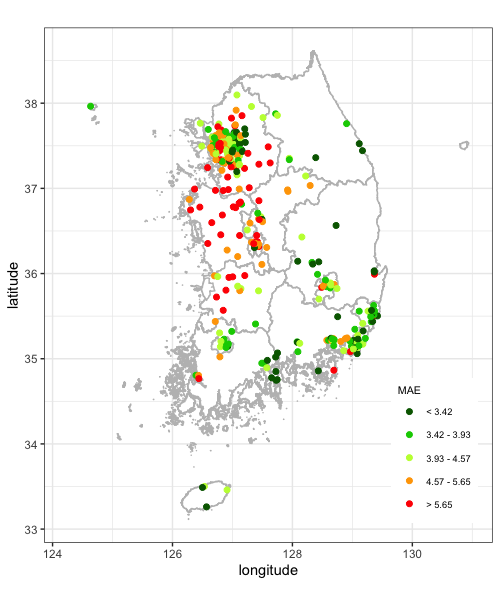}  
    \caption{Period 9, $h=1$.}
\end{subfigure}

\begin{subfigure}{0.32\textwidth}
    \centering
    \includegraphics[width=5cm]{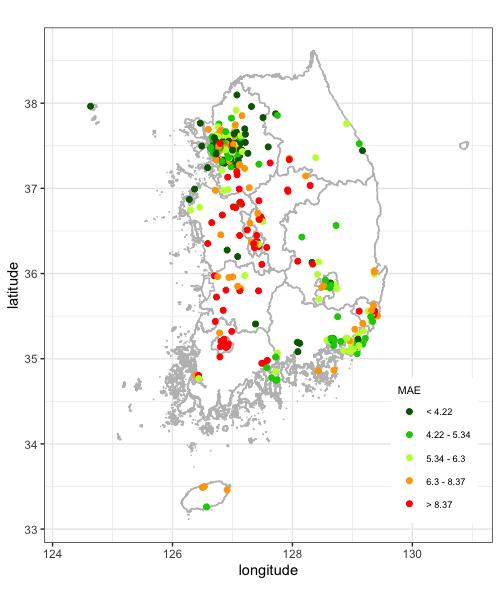}  
    \caption{Period 1, $h=6$.}
\end{subfigure}
\begin{subfigure}{0.32\textwidth}
    \centering
    \includegraphics[width=5cm]{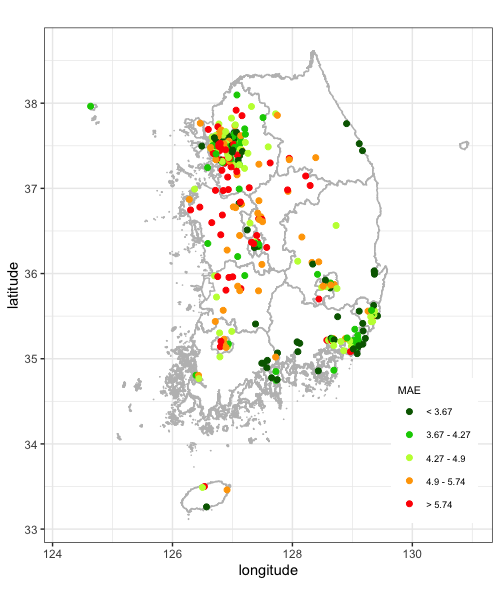}  
    \caption{Period 5, $h=6$.}
\end{subfigure}
\begin{subfigure}{0.32\textwidth}
    \centering
    \includegraphics[width=5cm]{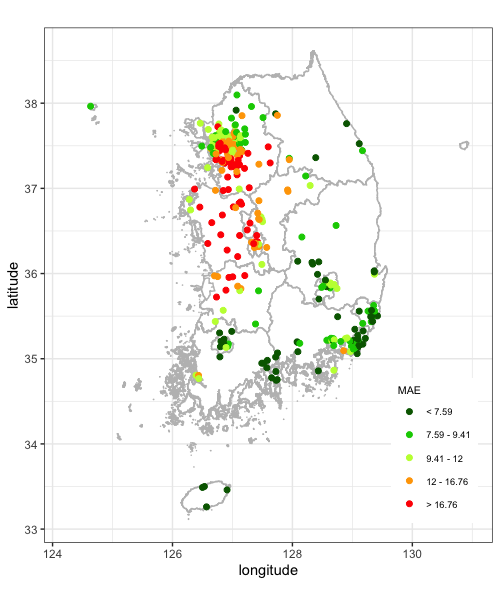}  
    \caption{Period 9, $h=6$.}
\end{subfigure}
\caption{MAE values of 308 stations in periods 1, 5, and 9 with forecasting horizon $h = 1, 6$. The five colors distinguish MAE values within five intervals separated by 0.2, 0.4, 0.6, and 0.8 quantiles of the MAE values.}
\label{fig:mae map}
\end{figure}

Finally, Figure \ref{fig:mae map} provides an overview of the forecasting performances across the stations using the proposed method. Panels (a)--(c) show the MAE values for each station during three specific periods, 1, 5, and 9, with a forecasting horizon of $h=1$. Panels (d)--(f) show the same information but when $h=6$. The MAE value for the $i$th station is computed as
    $MAE=\sum_{t=t_1}^{t_T} \big |\hat{y}_{t,i}-y_{t,i}\big |/T,~(T=48)$.
The five colors in the figure distinguish the MAE values within five intervals separated by 0.2, 0.4, 0.6, and 0.8 quantiles of the MAE values. The gaps across the periods or the values of $h$ are insignificant, but there is a noticeable tendency for more dense regions to have more accurate predictions than less dense regions. This can be explained by the fact that denser areas benefit from predictors with more relevant patterns and better use of information from nearby stations when estimating the weights.    

\begin{table}
   \caption{MAE results of periods 1--6. These are the results of $h$-step ahead forecasts by the five competing methods, the proposed method, and the extended method. MAE values are computed for each $h$ and test period. The lowest MAE is achieved by the proposed method or the extended method in all cases.}
   \small
   \centering
   {\renewcommand{\arraystretch}{1.1}
   \setlength{\tabcolsep}{3pt}
   \begin{tabular}{c|cccccc|cccccc}
   \toprule
   \hline
   $h$ & 1 & 2 & 3 & 4 & 5 & 6 & 1 & 2 & 3 & 4 & 5 & 6 \\
   \hline
  \multicolumn{1}{c}{} & \multicolumn{6}{c|}{Period 1} & \multicolumn{6}{c}{Period 2} \\
   \hline
Na\"ive & 3.55 & 4.84 & 5.80 & 6.52 & 7.12 & 7.60 & 4.17 & 5.95 & 7.15 & 8.11 & 8.91 & 9.52 \\
AR & 3.47 & 4.62 & 5.41 & 5.96 & 6.35 & 6.63 & 4.15 & 5.94 & 7.16 & 8.10 & 8.86 & 9.46 \\
ARIMA & 3.51 & 4.74 & 5.63 & 6.29 & 6.80 & 7.22 & 4.14 & 5.93 & 7.12 & 8.05 & 8.83 & 9.43 \\
Near stations & 3.42 & 4.62 & 5.48 & 6.10 & 6.61 & 7.06 & 3.99 & 5.64 & 6.74 & 7.58 & 8.31 & 8.96 \\
SW(2002b) & 3.47 & 4.70 & 5.56 & 6.11 & 6.48 & 6.83 & 4.04 & 5.62 & 6.65 & 7.48 & 8.27 & 9.04 \\
Proposed & \textbf{ 3.31 } & \textbf{ 4.29 } & 5.00 & 5.50 & 5.90 & 6.26 & \textbf{ 3.93 } & \textbf{ 5.44 } & \textbf{ 6.43 } & \textbf{ 7.22 } & \textbf{ 7.84 } & \textbf{ 8.30 } \\
Extended & 3.33 & 4.32 & \textbf{ 5.00 } & \textbf{ 5.46 } & \textbf{ 5.80 } & \textbf{ 6.15 } & 3.94 & 5.48 & 6.46 & 7.28 & 7.94 & 8.43 \\
\hline
\multicolumn{1}{c}{} & \multicolumn{6}{c|}{Period 3} & \multicolumn{6}{c}{Period 4} \\
   \hline
Na\"ive & 2.24 & 3.02 & 3.53 & 3.96 & 4.30 & 4.63 & 2.37 & 3.16 & 3.68 & 4.11 & 4.48 & 4.77 \\
AR & 2.30 & 3.11 & 3.68 & 4.16 & 4.57 & 4.94 & 2.34 & 3.05 & 3.51 & 3.87 & 4.14 & 4.36 \\
ARIMA & 2.26 & 3.02 & 3.56 & 3.99 & 4.35 & 4.68 & 2.33 & 3.07 & 3.55 & 3.94 & 4.26 & 4.51 \\
Near stations & 2.18 & 2.87 & 3.40 & 3.89 & 4.31 & 4.72 & 2.27 & 2.92 & 3.39 & 3.73 & 3.97 & 4.14 \\
SW(2002b) & 2.21 & 2.90 & 3.43 & 3.90 & 4.34 & 4.79 & 2.32 & 3.06 & 3.56 & 3.87 & 4.07 & 4.16 \\
Proposed & \textbf{ 2.10 } & \textbf{ 2.67 } & \textbf{ 3.03 } & \textbf{ 3.33 } & \textbf{ 3.57 } & \textbf{ 3.75 } & \textbf{ 2.20 } & \textbf{ 2.78 } & \textbf{ 3.18 } & \textbf{ 3.48 } & \textbf{ 3.70 } & \textbf{ 3.85 } \\
Extended & 2.10 & 2.69 & 3.06 & 3.36 & 3.59 & 3.77 & 2.22 & 2.81 & 3.20 & 3.49 & 3.70 & 3.87 \\
\hline
\multicolumn{1}{c}{} & \multicolumn{6}{c|}{Period 5} & \multicolumn{6}{c}{Period 6} \\
   \hline
Na\"ive & 2.64 & 3.65 & 4.38 & 5.00 & 5.50 & 5.92 & 2.38 & 3.21 & 3.77 & 4.23 & 4.63 & 4.96 \\
AR & 2.62 & 3.55 & 4.16 & 4.63 & 4.99 & 5.29 & 2.52 & 3.49 & 4.21 & 4.82 & 5.37 & 5.82 \\
ARIMA & 2.62 & 3.60 & 4.27 & 4.80 & 5.22 & 5.57 & 2.43 & 3.28 & 3.89 & 4.38 & 4.81 & 5.20 \\
Near stations & 2.46 & 3.31 & 3.96 & 4.47 & 4.87 & 5.19 & 2.37 & 3.20 & 3.87 & 4.42 & 4.88 & 5.15 \\
SW(2002b) & 2.52 & 3.44 & 4.14 & 4.68 & 5.09 & 5.43 & 2.34 & 3.10 & 3.70 & 4.17 & 4.45 & 4.54 \\
Proposed & 2.44 & 3.22 & 3.79 & 4.26 & \textbf{ 4.55 } & 4.73 & \textbf{ 2.22 } & \textbf{ 2.83 } & \textbf{ 3.27 } & \textbf{ 3.65 } & \textbf{ 3.98 } & \textbf{ 4.25 } \\
Extended & \textbf{ 2.43 } & \textbf{ 3.21 } & \textbf{ 3.79 } & \textbf{ 4.24 } & 4.55 & \textbf{ 4.71 } & 2.24 & 2.87 & 3.33 & 3.73 & 4.06 & 4.30 \\
   \hline
   \bottomrule
   \end{tabular}
   }
\label{table:result1}
\end{table}

\begin{table}
   \caption{MAE results of periods 7--12. These are the results of $h$-step ahead forecasts by the five competing methods, the proposed method, and the extended method. MAE values are computed for each $h$ and test period. The lowest MAE is achieved by the proposed method or the extended method in most cases, except for some $h$'s in periods 9 and 10.} 
   \small
   \centering
   {\renewcommand{\arraystretch}{1.1}
   \setlength{\tabcolsep}{3pt}
   \begin{tabular}{c|cccccc|cccccc}
   \toprule
   \hline
   $h$ & 1 & 2 & 3 & 4 & 5 & 6 & 1 & 2 & 3 & 4 & 5 & 6 \\
   \hline
  \multicolumn{1}{c}{} & \multicolumn{6}{c|}{Period 7} & \multicolumn{6}{c}{Period 8} \\
   \hline
Na\"ive & 4.28 & 6.48 & 8.16 & 9.40 & 10.37 & 11.03 & 3.28 & 4.76 & 5.95 & 6.95 & 7.79 & 8.49 \\
AR & 4.13 & 6.10 & 7.42 & 8.28 & 8.89 & 9.24 & 3.19 & 4.49 & 5.43 & 6.14 & 6.68 & 7.07 \\
ARIMA & 4.19 & 6.29 & 7.78 & 8.83 & 9.62 & 10.16 & 3.20 & 4.53 & 5.51 & 6.27 & 6.86 & 7.31 \\
Near stations & 3.84 & 5.67 & 7.22 & 8.42 & 9.32 & 9.93 & 3.00 & 4.16 & 5.06 & 5.78 & 6.35 & 6.79 \\
SW(2002b) & 3.95 & 5.69 & 6.97 & 7.83 & 8.50 & 9.16 & 2.95 & 4.03 & 4.89 & 5.53 & 6.01 & 6.35 \\
Proposed & 3.79 & 5.48 & 6.67 & 7.56 & 8.17 & 8.52 & 2.95 & 4.01 & 4.91 & 5.66 & 6.23 & 6.57 \\
Extended & \textbf{ 3.76 } & \textbf{ 5.37 } & \textbf{ 6.44 } & \textbf{ 7.16 } & \textbf{ 7.68 } & \textbf{ 8.02 } & \textbf{ 2.91 } & \textbf{ 3.94 } & \textbf{ 4.75 } & \textbf{ 5.43 } & \textbf{ 5.92 } & \textbf{ 6.21 } \\
\hline
\multicolumn{1}{c}{} & \multicolumn{6}{c|}{Period 9} & \multicolumn{6}{c}{Period 10} \\
   \hline
Na\"ive & 4.74 & 6.95 & 8.57 & 9.92 & 11.03 & 11.98 & 3.25 & 4.97 & 6.34 & 7.55 & 8.60 & 9.54 \\
AR & 4.64 & 6.86 & 8.46 & 9.80 & 10.90 & 11.84 & 3.23 & 4.87 & 6.13 & 7.15 & 7.99 & 8.70 \\
ARIMA & 4.64 & 6.82 & 8.42 & \textbf{ 9.74 } & \textbf{ 10.84 } & \textbf{ 11.77 } & 3.26 & 4.98 & 6.32 & 7.43 & 8.37 & 9.16 \\
Near stations & 4.66 & 7.05 & 8.95 & 10.55 & 11.87 & 12.89 & 2.91 & 4.34 & 5.54 & 6.64 & 7.64 & 8.51 \\
SW(2002b) & 4.69 & 7.22 & 9.25 & 11.01 & 12.57 & 13.83 & 2.90 & 4.17 & \textbf{ 5.16 } & \textbf{ 6.01 } & \textbf{ 6.79 } & \textbf{ 7.38 } \\
Proposed & 4.55 & 6.72 & 8.36 & 9.86 & 11.11 & 12.08 & 2.85 & 4.14 & 5.24 & 6.19 & 6.97 & 7.60 \\
Extended & \textbf{ 4.55 } & \textbf{ 6.68 } & \textbf{ 8.34 } & 9.79 & 11.00 & 11.97 & \textbf{ 2.84 } & \textbf{ 4.12 } & 5.17 & 6.06 & 6.80 & 7.38 \\
\hline
\multicolumn{1}{c}{} & \multicolumn{6}{c|}{Period 11} & \multicolumn{6}{c}{Period 12} \\
   \hline
Na\"ive & 3.47 & 4.80 & 5.77 & 6.51 & 7.20 & 7.80 & 2.74 & 3.57 & 4.14 & 4.61 & 5.01 & 5.37 \\
AR & 3.44 & 4.72 & 5.59 & 6.23 & 6.78 & 7.22 & 2.71 & 3.47 & 3.98 & 4.37 & 4.68 & 4.94 \\
ARIMA & 3.44 & 4.72 & 5.63 & 6.30 & 6.91 & 7.42 & 2.70 & 3.48 & 4.00 & 4.40 & 4.73 & 5.02 \\
Near stations & 3.19 & 4.30 & 5.18 & 5.86 & 6.43 & 6.88 & 2.58 & 3.25 & 3.71 & 4.15 & 4.50 & 4.82 \\
SW(2002b) & 3.14 & 4.11 & 4.76 & 5.24 & 5.63 & 5.91 & 2.59 & 3.26 & 3.71 & 4.10 & 4.43 & 4.70 \\
Proposed & \textbf{ 3.10 } & 4.02 & 4.69 & 5.17 & 5.55 & 5.80 & \textbf{ 2.51 } & \textbf{ 3.09 } & 3.48 & 3.74 & \textbf{ 3.98 } & 4.21 \\
Extended & 3.10 & \textbf{ 4.00 } & \textbf{ 4.63 } & \textbf{ 5.06 } & \textbf{ 5.34 } & \textbf{ 5.55 } & 2.52 & 3.10 & \textbf{ 3.46 } & \textbf{ 3.73 } & 3.99 & \textbf{ 4.21 } \\
   \hline
   \bottomrule
   \end{tabular}
   }
\label{table:result2}
\end{table}

\section{Concluding remarks} \label{sec:Concluding remarks} 
In this study, we proposed a new method for forecasting with many predictors under various data distributions. This novel approach combines a forecasting method using a factor model with a quantile approach. We found that the existing method by \cite{stock2002forecasting}, which uses the approximate factor-based forecasting model, was not optimal for non-Gaussian data. In addition, the simulation study has shown that it performs much worse than time-series models in some scenarios when the data follow a heavy-tailed or highly skewed distribution. To address this issue, our proposed method incorporated quantile regression, which, in contrast to the OLS estimator, provides better estimates in terms of robustness, as evidenced by the bounded influential function. To fit the quantile regression, we utilized a quantile factor model, which extracts quantile factors that describe particular quantiles of the data instead of the mean factor. This approach successfully predicted specific quantiles of the target variable. The predicted quantiles at different levels between 0 and 1 were then aggregated using a Markov chain with weights based on the past trends in the target variable. The performance of the proposed method was finally demonstrated through a simulation study and real data analysis of $\text{PM}_{2.5}$ data in South Korea. In most cases, it was found to have the lowest mean absolute error in forecasting compared to other methods. 

This study emphasizes the importance of incorporating quantile approaches when forecasting non-Gaussian data and demonstrates the effectiveness of the proposed method. We believe that the proposed method holds significant potential for applications beyond environmental forecasting tasks, extending its usage to various fields dealing with large data panels, such as economics, politics, and sociology. Due to its generality in covering a broad range of data distributions, it can be applied to various datasets.

To further improve the proposed method, future research can explore how to choose the grid of quantile levels. In this study, we arbitrarily chose five levels: 0.1, 0.3, 0.5, 0.7, and 0.9, but other combinations may yield better forecasts. For example, we can choose the quantile levels based on domain knowledge of data, which might suggest the important quantiles of the data with unique characteristics. Alternatively, it can be explored based on the underlying distributional characteristics of the data. In addition, the proposed method can be improved by the appropriate number of factors when fitting quantile factor models. According to \cite{bai2002determining}, there are many possible penalty terms $g(n,T)$ for the information criteria, and the estimations are inconsistent depending on the choice of $g(n,T)$. Therefore, an advanced way of selecting the penalty term may be needed. These are reserved for future research. 

\section*{Acknowledgments} 
This research was supported by the National Research Foundation of Korea (NRF) funded by the Korea government (2021R1A2C1091357). 

\bibliographystyle{apalike}
\bibliography{pmforecast}

\end{document}